\documentclass[epjST]{svjour}
\usepackage{amssymb}
\usepackage{amsbsy}
\usepackage{amsmath}
\usepackage{graphicx}
\usepackage{amsmath}
\usepackage{times}
\usepackage{color}
\usepackage{subfigure}
\usepackage{setspace}
\usepackage{bm}% bold math
\newcommand\bea{\begin{eqnarray}}
\newcommand\eea{\end{eqnarray}}
\newcommand\beq{\begin{equation}}
\newcommand\eeq{\end{equation}}
\newcommand\bib{\bibitem}

\newcommand{\prl}{Phys. Rev. Lett.}

\begin{document}

\title{ Controlling quantum critical dynamics of isolated systems}
\author{A. del Campo$^{1}$ and K. Sengupta$^{2}$}

\institute{$^1$ Department of Physics, University of Massachusetts
Boston, Boston, MA 02125, USA.
\\$^2$ Theoretical Physics Department, Indian Association for the
Cultivation of Science, Jadavpur, Kolkata-700032, India.}

\date{\today}

%\begin{abstract}
\abstract{ Controlling the non adiabatic dynamics of isolated
quantum systems driven through a critical point is of interest in a
variety of fields ranging from quantum simulation to finite-time
thermodynamics. We briefly review the different methods for
designing protocols which minimize excitation (defect) production in
a closed quantum critical system driven out of equilibrium. We chart
out the role of specific driving schemes for this procedure, point
out their experimental relevance, and discuss their implementation
in the context of ultracold atom and spin systems.}
%\end{abstract}

%\pacs{73.43.Nq, 05.70.Jk, 64.60.Ht, 75.10.Jm}

\maketitle

\section{Introduction}
\label{intro}

The physics of closed quantum systems driven out of equilibrium has
received a lot of theoretical and experimental attention in recent
years. One of the central issues in this field involves
understanding excitation or defect production resulting from a
driving protocol. The associated dynamics becomes specially
important when it involves the crossing of a quantum critical point
 \cite{pol1,dziar1,dutta1,DZ14,EFG14}. The breakdown of adiabatic
dynamics is often characterized by the probability of not ending up
in the ground state which can be quantified using the density of
excitations $n$. It is as well captured by the excess of energy over
the instantaneous ground state $E_0$,
$Q=\langle\psi|H|\psi\rangle-E_0$. Many other quantities can be
used,  including the density of topological defects
  \cite{Kibble76,Zurek96} (related to, but different
from $n$  \cite{Dziarmaga05,USF10}), the fidelity, etc. It is well
known that when the crossing of a quantum critical point is induced
by a slow linear quench of a parameter of the system Hamiltonian at
a given rate $\omega$, a universal scaling law is observed. For
instance, the excitation  density $n$ and the residual energy $Q$
exhibit a power-law behavior
\begin{eqnarray}
n \sim \omega^{d \nu/(zv+1)},\quad Q\sim \omega^{(d+z)\nu/(z\nu
+1)},\label{eq:scalinglaws}
\end{eqnarray}
where $d$ is the dimension of the system and $z$ and $\nu$ are the
dynamic  and correlation length critical exponents
 \cite{Kibble76,Zurek96,Damski05,ZDZ05,Dziarmaga05,pol2}. Recent
findings suggest that analogous signatures of universality are still
present in dynamics of strongly-coupled systems, e.g., described by
holographic duality  \cite{SDZ14,Chesler14,ks0}. Such scaling laws
can also be extended to cases where the system passes through a
critical surface  \cite{ks1}, for non-linear ramps
 \cite{BP08,ks2,pol3,deGrandi10,sondhi1} and in the presence of an external
control parameter with a self-consistent dynamics   \cite{KKP14},
and indicate an inevitable growth of $n$ with increasing $\omega$.
The root of such an increase owes its existence to the very nature
of the critical point; as the characteristic energy gap $\Delta$
closes in its neighborhood, no drive can remain  nearly adiabatic
and the Landau criterion for excitation production $d\Delta/dt \ge
\Delta^2$ is always satisfied
  \cite{Damski05}.

Such an increase of $n$ and $Q$ is, however, disadvantageous for the
purpose of quantum computation, quantum state preparation,  the
control of non adiabatic processes in quantum critical system, and
that of the associated work fluctuations,  of interest to optimize
the efficiency of quantum devices operating at the nanoscale. In all
these cases, it is necessary to implement dynamical protocols which
take a quantum system from one state to another in a finite amount
of time. To this end, a quantum system is typically prepared in its
ground state $|\psi_0\rangle$ for
 given values of the control parameters  $\{\lambda_i\}$ of the system Hamiltonian
$H_0[\{\lambda_i\}]$. Subsequently, a dynamics is induced by either
changing these parameters as a function of time $\{\lambda_i \equiv
\lambda_i(t)\}$, or by subjecting the system to an external possibly
time-dependent perturbation $H_1(t)$ for a  pre-determined
finite amount of time $T$. The resulting quantum evolution, governed
by the time-dependent Schr\"odinger equation $i \hbar \partial_t
|\psi(t)\rangle = [H_0(t)+H_1(t)] |\psi(t)\rangle$, takes the system to a new
state $|\psi_f\rangle \equiv |\psi(T)\rangle$ at the end of the
process. The question arises as to how a specific final quantum
state can be reached $|n\rangle$ at the end of the evolution with
close to unit fidelity,  i.e., ensuring  that $|\langle
n|\psi_f\rangle|^2 \simeq 1$. If $|n\rangle$ happens to be the
ground state of the final Hamiltonian, a unit fidelity can be
achieved via a reduction of the formation of excitations during the
dynamics, or what appears to be more challenging, by canceling
excitations in a non adiabatic protocol upon completion of the
process. We shall see that a variety of shortcuts to adiabaticity in
critical systems achieved the latter goal.

At this stage, let us clarify that we shall focus exclusively on  driven systems 
which obey unitary dynamics and that are described by a time-dependent Hamiltonian. Hence, we consider systems decoupled from the surrounding environment, 
up to the set of external control parameters. Consequently, we shall not dwell on the
possible use of the environment to reduce excitation formation
\cite{Patane08,Patane09}, the design of open quantum dynamics
\cite{Vacanti14}, or the use of Hamiltonian quantum controls to
effectively decouple the system from its environment
 \cite{VS98,VKS99,SGB13}. We note that even within this restricted
territory, efforts to guide the non adiabatic dynamics of driven
quantum systems while mimicking adiabaticity have already led to a
broad variety of theoretical and experimental results scattered in
the literature  \cite{DR03,Berry09,Chen10,Torrontegui13,Jarzynski13,DZ14,OK14}, with
applications to quantum fluids
 \cite{Muga09,Schaff1,Schaff2,Stefanatos10,delcampo11,RC11,DB12,polls1,SRS13,delcampo13,Rohringer13},
trapped ions  \cite{Torrontegui11,ions1,ions2,Masuda12,Palmero13,DJD14} and
effective few-level  systems \cite{DR03,Berry09,expCD1,Chen10b,expCD2,Hegerfeldt13}.
%Any form is good, here a small twist
In this review, we shall focus on an prominent sub-area,
namely, the control of  quantum critical dynamics
 \cite{KV97,DLZ99,BP08,Zurek09,pol3,Caneva09,dc1,dc2,kstr1,kstr2}.
The purpose of the present review is to briefly outline some recent
developments aimed at tailoring, controlling and reducing excitation
formation in driven quantum critical systems and discuss the
feasibility of their implementation in realistic experimental
systems.

A natural strategy  to warrant an excitation-free evolution in
finite-size gapped systems is to comply with the adiabatic theorem
\cite{JRS07,MC04}. Excitation formation is suppressed whenever the quench time is longer than 
the inverse of a given power of the minimum energy gap, 
that can be efficiently computed, see e. g., \cite{Mandra14}.

Whenever the critical point is precisely known
and an exquisite control of the external parameter is available,
knowledge of the scaling laws in (\ref{eq:scalinglaws}) can be used to design optimal
nonlinear quenches  for which the excitation density is reduced
given a fixed duration of the process  \cite{BP08,kstr2,dziar1}. One
can also resort on a more general, yet smooth time-dependence of the
external control  \cite{Lidar09}.

As an alternative, in spatially extended systems with finite range
interactions one can implement an inhomogeneous driving. In such scenario, 
criticality is first reached locally in a finite region of the system, whose 
spatial extent grows subsequently with time. Tuning the
velocity  of the critical front paves the way to a fully adiabatic
crossing of the phase transition whenever its value does not surpass
the second sound velocity \cite{KV97,DLZ99}. This idea was
introduced in a classical setting
 \cite{KV97,DLZ99,Zurek09,delcampo10,DRP11,WDGR11} and has been
experimentally explored  in the context of kink formation in trapped
ion chains  \cite{Schaetz13,EH13,Ulm13,Tanja13} and soliton creation
in harmonically confined Bose-Einstein condensates
 \cite{Lamporesi13}. However,  its key  tenets have been shown to
hold in quantum systems as well  \cite{ZD08,DZ09,DM10,DM10b,CK10}
(see Ref.\ \cite{dc2} for an updated account).

In what follows,  we shall focus on
 three techniques to mimic adiabaticity. The first of these is generally referred to as the counterdiabatic driving technique and
was introduced by Demirplak and Rice  \cite{DR03}, and elaborated by
Berry  \cite{Berry09} (see also Ref.\ \cite{ASY87}). It involves a
modification of the system Hamiltonian $H_0(t)$ by a suitably chosen
auxiliary term $H_1(t)$. In this protocol, $H_1(t)$ is chosen so
that the adiabatic approximation to $H_0(t)$ becomes the exact
solution of the many-body time dependent Schr\"odinger equation with
$H(t)=H_0(t)+H_1(t)$  \cite{DR03,Berry09}. Such a procedure has been
studied for several systems including a large variety if
single-particle, many-body and nonlinear systems
\cite{DR03,Berry09,dc1,delcampo13,DJD14,polls1}. In quantum critical
systems, its experimental implementation is expected to be
complicated as the auxiliary term is generally  nonlocal and include
many-body interactions, although its form can be suitably
tailored under given resources  \cite{OK14,Saberi14}.
 The second class of
methods involve designing an optimal protocol that leads to maximal
reduction of excitations for a fixed evolution time $T$
 \cite{pol3,Caneva09,RC11}. Optimal protocols can be difficult to find
for complicated interacting quantum systems; however, they have been
computed for specific many-body models  \cite{Caneva09,RC11}. Finally,
the third approach involves simultaneous variation of two parameters
of a system Hamiltonian  \cite{kstr1}. The
first of these controls the proximity of the quantum system to the
critical point while the second determines the phase space available for  excitation production.
This technique also applies to
experimentally realizable non-integrable model  \cite{kstr2}. While
this method does not constitute an optimal protocol,
its main advantage is the possibility relatively simple
experimental implementation.

The rest of this article is organized as follows.
We devote Sec.\
\ref{secCD}  to counterdiabatic driving,
Sec.\ \ref{octh1} to the applications of the
optimal control in excitation suppression and
Sec.\ \ref{sau1} to the two-rate protocol.
We close with a discussion in Sec.\ \ref{conc}.

%a section to each of these techniques and close with a discussion in Sec.\ \ref{conc}.
%After devoting the next three sections to
%In Sec.\
%\ref{secCD}, we discuss the details of the first of these protocols
%and some specific context in which they have been applied. This is
%followed by Sec.\ \ref{octh1}, where we discuss some aspects of the
%optimal control techniques for defect reduction in closed quantum
%system. Next, in Sec.\ \ref{sau1}, we discuss the two-rate protocol,
%where two parameters of a Hamiltonian is simultaneously changed.
%Finally, we conclude in Sec.\ \ref{conc}.
%,where two parameters of a Hamiltonian is simultaneously changed.

\section{Counterdiabatic driving}
\label{secCD}

In this section, we  review the counterdiabatic driving technique
also known as  transitionless quantum driving  \cite{DR03,Berry09}. The key idea behind this
approach is to find an auxiliary counterdiabatic Hamiltonian
$H_1(t)$, which when added to the system Hamiltonian $H_0 (t)$
ensures that the adiabatic approximation to $H_0 (t)$ becomes the
exact solution to the dynamics generated by $H(t)=H_0(t)+H_1(t)$, even in
the absence of slow driving. As a result, $H(t)$ drives a ``
fast-motion video'' of the adiabatic dynamics associated with $H_0
(t)$. The form of $H$ corresponding to a given $H_0$ can be
obtained as follows. Consider the Schr\"odinger
equation
\begin{eqnarray}
H_0(t) |n\rangle &=& E_n(t) |n\rangle,  \label{inst}
\end{eqnarray}
where $\{E_n(t)\}$ and $\{|n(t)\rangle\}$ denote the set of
instantaneous eigenvalues and eigenstates, respectively. Next,
select as a  target trajectory,
\begin{eqnarray}
|\psi_n\rangle = e^{-\int_0^t dt'[iE_n(t')/\hbar +\langle n|\partial_{t'}
n\rangle]} |n\rangle, \label{psi1}
\end{eqnarray}
i.e., the adiabatic evolution of $|n\rangle$ which would
only describe the dynamics generated by $H_0 (t)$ under slow
driving. Note that the phase factor in Eq. (\ref{psi1}) incorporates
the dynamic contribution as well as the geometric phase, generated
by the Berry potential $ A_n(t')= i\langle n|\partial_{t'}
n\rangle$. One then needs to design a Hamiltonian $H(t)$ which
satisfies
\begin{eqnarray}
i \hbar \partial_t |\psi_n\rangle &=& H(t) |\psi_n\rangle =
[H_0(t)+H_1(t)] |\psi_n\rangle, \label{psi2}
\end{eqnarray}
so that $|\psi_n(t)\rangle$ remains the instantaneous ground state of
$H_0$ up to a phase. The spectral decomposition of the time-evolution operator simply reads
$U(t,0)=\sum_n |\psi_n(t)\rangle\langle n(0)|$, from which the required Hamiltonian can be derived using the identity
\begin{eqnarray}
H(t)=i\hbar (\partial_tU(t,0))U(t,0)^{\dag}.
\end{eqnarray}
It follows that
\begin{eqnarray}
H(t) &=& \sum_n E_n |n\rangle \langle n|  + \sum_n
[i \hbar |\partial_t n\rangle\langle n| -\hbar A_n(t) |n\rangle \langle n|],
\label{hexp}\nonumber\\
\end{eqnarray}
where the first and second terms on the RHS can be recognized as the system Hamiltonian and the auxiliary term, respectively.
The latter can be rewritten as
\begin{eqnarray}
H_1(t) &=& i\hbar \sum_{m\ne n}\sum_n \frac{ \langle m|\partial_t
H_0(t)|n\rangle}{E_n-E_m} |m\rangle\langle n|,
\end{eqnarray}
where $|m\rangle, |n\rangle$ are the instantaneous eigenstates of
$H_0$ and we have assumed that the system is non-degenerate. As
expected $H_1(t)$ vanishes in the truly adiabatic limit while its
norm increases with the rate of change of the system Hamiltonian
along  \cite{DR03,dc1}. We note that 
computing  $H_1$ requires full knowledge of the instantaneous
spectral properties of the system, i.e., that of  $\{E_n(t)\}$ and
$\{|n(t)\rangle\}$. For a moderate system size this information can
be accessed via numerical methods, while for arbitrary system size
its determination becomes a theoretically challenge  beyond simple
integrable models. 

This requirement can be removed  by introducing hybrid methods 
relying on a combination of counterdiabatic driving and optimal control \cite{LGM2}.
Indeed, a practical advantage of counterdiabatic driving 
lies in its power to determine
 approximate expressions of $H_1$ leading to a controlled reduction, instead of a
complete suppression, of excitation production during  critical dynamics
 \cite{dc1,Saberi14,LGM2}. To this end, one
can further exploit the freedom associated with the choice of the
phase of $|\psi_n\rangle$, which need not include the dynamic and
geometric contributions. This is, given a unitary $G(t)$, the time
evolution along the trajectory $G(t)|\psi_n\rangle$ is generated by
a unitarily equivalent Hamiltonian
\begin{eqnarray}
H_G=GHG^{\dag}-i\hbar G\partial_tG^{\dag},
\end{eqnarray}
whose physical properties are normally completely different from
those of $H$  \cite{bookinv}. One can allow for excitations  to occur along the
dynamics and impose boundary conditions at the beginning and end of
the protocol so that $G(t)$ reduces then to the identity. This
approach has proven extremely useful in designing
experimentally-realizable shortcuts to adiabaticity
 \cite{Berry09,Ibanez12,delcampo13,DJD14} as demonstrated in quantum
optical systems  \cite{expCD1,expCD2} and low-dimensional quantum
gases  \cite{Rohringer13}, while realizations in critical systems
remain to be explored.

Before considering the dynamics across critical point,
it is illuminating to understand the counterdiabatic driving
of single-particle problems.
A specific instance of such a problem which is of direct relevance
to many-body systems is the time-dependent two-level system whose
Hamiltonian is given by
\begin{eqnarray}
H_0(t) &=& \lambda(t) \sigma_z + \Delta \sigma_x,
\end{eqnarray}
with instantaneous eigenenergies $E_{\pm}(t) = \pm
\sqrt{\lambda^2(t) +\Delta^2}$. The avoided level crossing for such
a system takes place at $\lambda(t)=0$. Then, a straightforward
calculation  leads to \cite{DR03}
\begin{eqnarray}
H_1(t) = -\frac{\Delta \partial_t
\lambda(t)}{2(\lambda^2(t)+\Delta^2)} \sigma_y. \label{tl1}
\end{eqnarray}
This result can be directly applied to a class of integrable models known as quasi-free fermion systems
that includes paradigmatic models in statistical mechanics such as the Ising and the XY models in $d=1$ and
the Kitaev model in $d=2$  \cite{dc1}.
In what follows we are going to discuss
the Ising model explicitly. The Hamiltonian of the Ising model for
$d=1$ is given by
\begin{eqnarray}
H_{0} &=& -\sum_{\langle ij\rangle} S_i^z S_j^z + g(t)
\sum_i S_i^x,
\end{eqnarray}
where $g(t)$ denotes a time-dependent dimensionless transverse
magnetic field, $S_i^{x,z}$ denotes the spin operators on the
$i^{\rm th}$ site of the 1D lattice, and $\langle ij\rangle$
indicates that $i$ and $j$ are neighboring sites. Such a Hamiltonian
can be mapped, via well-known Jordan-Wigner transformation, to a set
of fermionic two-level systems whose Hamiltonian is given by
$H_0=\sum_{k>0} \psi_k^{\dagger} H_k \psi_k$, where $\psi(k)= (c_k
,c_{-k}^{\dagger})$ are two component Fermion operators and $H_k$ is
given by
\begin{eqnarray}
H_k &=& 2[g(t)-\cos(k)]\sigma_z + \sin(k) \sigma_x.
\end{eqnarray}
Using Eq. (\ref{tl1}), one can thus write an expression for $H_1(t)$
 \cite{dc1}
\begin{eqnarray}
H_1 &=& -\sum_{k>0} \frac{\sin(k) \partial_t g(t)}{2[1+g^2(t)-2g(t)
\cos(k)]} \psi_k^{\dagger} \sigma_y \psi_k.
\end{eqnarray}
\begin{figure}[tbp]
\begin{center}
\includegraphics[width=0.5\textwidth]{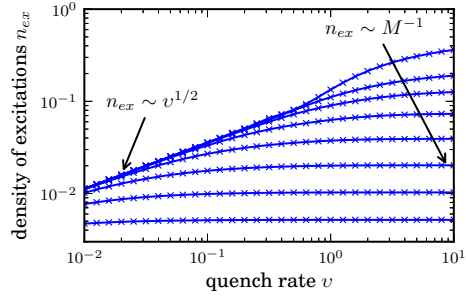}
\end{center}
\caption{(Color online) Reduction of excitation density in a 1D
Ising chain quench through its critical point by a ramp of the
transverse field. The exact auxiliary counterdiabatic driving term $H_1$
involves  multi-body interactions which extend over the  whole
system. A direct real-space truncation of $H_1$ effectively
suppresses excitations with wave vector $k>1/M$, whenever $H_1$
includes interactions of up to
 $M$-spins. The value of $M$ is chosen to be
$0$, $1$, $2$, $4$, $8$, $16$, $32$ and $64$ from top to bottom. The
figure demonstrates a gradual suppression of  excitation density as the
cutoff $M$ is increased. Taken from \cite{dc1}.}\label{fig1}
\end{figure}
The neat form of $H_1$ in  the Fermion language is not preserved in spin-space  \cite{dc1}.
Indeed,  a reverse Jordan-Wigner
transformation reveals that $H_1$ involves
multiple spin terms. Explicitly, for an even number of spins  under periodic
boundary conditions  $H_1$ takes the form
\begin{eqnarray}
H_1=-\frac{dg}{dt}\sum_{m=1}^{N/2} h_m(g) H_1^{[m]},
\end{eqnarray}
where $\{h_m(g)\}$ are real coefficients decaying over the
equilibrium correlation length and $H_1^{[m]}$ involves $m$-body
interaction extended over $m$ adjoint spins. The efficiency of an
approximate expression obtained by direct truncation in spin-space
restricting the sum to $M<N/2$  is shown in Fig.\ \ref{fig1} and it
clearly demonstrates a reduction of the excitation density as
$M$ is increased. Similarly, the  general counterdiabatic driving
term for quasi-free fermion systems was presented in ref.\
 \cite{dc1}, and is generally expected to be nonlocal. The
reader is referred to Ref.\ \cite{Takahashi13} for a detailed
discussion of counterdiabatic driving in  the XY model and the
Lipkin-Meshkov-Glick model in the thermodynamic limit. 
However,  for the typical number of spins of relevance to current
efforts in quantum simulation, finite size corrections play a key
role and need  to be taken into account as pointed out in
\cite{Damski14,LGM2}.
%value of the coefficients $h_m(g)=\frac{1}{2N}\sum_k \frac{\sin(k)\sin(mk)}{g^2-2g\cos(k)+1}$  \cite{DamskiTBP}.

As an alternative to the direct real-space truncation of $H_1$, one
can adopt a practical approach and look for an
approximate auxiliary term $\tilde{H}_1=\sum_k\alpha_kL_k$ realizable in terms of
the set   $\{L_k\}$ of available controls. The optimal value of the
coefficients $\alpha_k$ can be determined by a variational principle
of the form
 \cite{OK14,Saberi14}
\begin{eqnarray}
\min_{\{\alpha_k(t)\}}\||(H_1 -\tilde{H}_1) |\psi_{n}(t)\rangle ||^2.
\end{eqnarray}
Tailoring the counterdiabatic auxiliary interactions in this way, it
was shown that few-body short-range interactions suffice to generate
an effectively adiabatic dynamics.

We shall close this section pointing out that the experimental
implementations of counterdiabatic driving scheme in many-body
systems could be pursued using stroboscopic techniques
 \cite{Muller11} for digital quantum simulation with either trapped
ions  \cite{Barreiro11,Casanova12} or polar molecules
 \cite{Weimer10}.

\section{Optimal Control Methods}
\label{octh1}

 The method of optimal control exploits variational calculus to determine a driving
protocol which minimizes a given cost function, e.g.,  the density of excitations.  A variety of methods have
been proposed in the literature  \cite{oc1}.

In the context of quantum critical dynamics, the determination of the optimal non-linear ramp driving a phase transition  was discussed in Ref.\ \cite{pol3}. Consider a modulation of
a Hamiltonian parameter $g$ described by $g(t) =g_0|t/T|^r$  during
the  interval $[-T,T]$, according to which the critical point  is
crossed at $t=0$. It can be shown  \cite{pol3} that the optimal
protocol for minimization of defect production occurs when the
exponent $r$ is chosen to be
\begin{eqnarray}
r = -(z\nu)^{-1} \ln [\delta C^{-1}\ln(C/\delta)], \label{optpower}
\end{eqnarray}
where $C$ is a non-universal  system-specific constant of the order unity, $z$ and $\nu$ are the dynamic 
and correlation length critical exponents,
$\delta=1/(T\Delta_0)$, and $\Delta_0$ is a typical low-energy scale
in the system for $g=g_0$  \cite{pol3}. This non-trivial optimal power for $r$
is a function of the passage time but is independent of the dimension of
the system. The defect density $n_{\rm opt}$
produced during such a drive scales as
\begin{eqnarray}
n_{\rm opt} \simeq [\delta C^{-1} \ln(C/\delta)]^{d/z}
\end{eqnarray}
and satisfies $n_{\rm opt} \ll n_{\rm lin}$, where $n_{\rm lin}$ is
the defect density generated due to a linear ramp ($r=1$).

The above analysis relies on universal dynamics of phase transitions.
A more systematic approach to  suppress excitations is based on recasting the minimization
problem as a standard problem in
optimal control theory  \cite{oc1}, a strategy explored in a recent series of works
 \cite{MT09,Caneva09,RC11,DCM11,CCM11,WNR14}.  Assume a
quantum system with a Hamiltonian of the form $H_0= \sum_{i=1, N} \lambda_i h_i$
where $h_i$ are local operators with dimensions of energy and
$\lambda_i$ are the  corresponding dimensionless couplings. Let
the system be initialized in its ground state $|\psi_1\rangle$ for
$\{\lambda_i\}= \{ \lambda_0 \}$ and consider a modulation of the system Hamiltonian such that 
$\{\lambda_i(T)\}=\{\lambda_f\}$, with $|\psi_2\rangle$ being the
ground state of $H(\{\lambda_f\})$. Denoting the state of the system at the end of the evolution is $|\psi_T\rangle$, one looks for an optimal time-dependence of $\{\lambda\}$ which
maximizes the overlap $|\langle \psi_T|\psi_2\rangle|^2$. As
expected, the result turns out to be system
specific and to date, solutions are known for a few model system
only  \cite{Caneva09,RC11}. In particular, determining the optimal protocol in high
dimensional ($d>1$) non-integrable interacting quantum many-body
Hamiltonians constitutes an important open problem in the field.

To illustrate the method, we follow ref.\
 \cite{RC11}, and choose the Luttinger liquid (LL) as the
specific system at hand. The Hamiltonian of the LL in the bosonic
representation reads
\begin{eqnarray}
H_{\rm LL} &=& u \sum_k [K \Pi_{k}\Pi_{-k} + K^{-1} k^2 \phi_k\phi_{-k}],
\label{llham}
\end{eqnarray}
where $u$ and $K$ are the velocity of the charge carriers and the Luttinger parameter,
respectively.
Here, $\phi_k$ denotes a bosonic field with momentum $k$ and
$\Pi_{k}=-i\partial_{\phi_k}$ is the conjugate momentum operator.
We note that $H_{\rm LL}$ is the low energy representation of the 1D
Hubbard model on a lattice
\begin{eqnarray}
H_{\rm Hubbard} = \sum_i \left[-(c_i^{\dagger} c_{i+1} + {\rm h.c})
+ V {\hat n}_i {\hat n}_{i+1}\right],
\end{eqnarray}
where $i$ denotes lattice coordinate, $c_i$ is the annihilation
operator for the bosons at site $i$, and ${\hat n}_i =c_i^{\dagger}
c_i$ is the fermion number operator. The LL description of the
low-energy sector of this model holds for $-2<V<2$ where the system
is gapless; for $|V|>2$, a charge-density wave (CDW) gap opens up.
The parameter $V$, in the LL regime, can be related to $K$ and $u$
directly via Bethe Ansatz, see e.g., ref.  \cite{Cazalilla04}.

Let us consider the system to be in its critical point at $V=2$ and
study the dynamics induced by a change of $V$, or equivalently,  $u$ and $K$. We shall
assume that this dynamics can be described in terms of LL
Hamiltonian with $u\equiv u(t)$ and $K\equiv K(t)$.
Using $\Pi_k = -i \partial_{\phi_k}$ and solving the
Schr\"odinger equation for the many-body wavefunction $|\psi\rangle=
\prod_k |\psi_k\rangle$, it follows that
\begin{eqnarray} \{i \partial_t - u(t) [-K(t)/4 (\partial_k^2)
+k^2 \phi_k^2 /K(t) ] \} |\psi_k\rangle =0,
\end{eqnarray}
which has a straightforward solution for $\psi_k = \langle
k|\psi_k\rangle$:
\begin{eqnarray}
\psi_k &=& (2k {\rm Re}[z_k(t)]/\pi)^{1/4} e^{-k z_k(t)
\phi_k^2}, \nonumber\\
i \partial_t z_k(t) &=& k u(t)K(t)[z_k(t)^2-1/K^2(t)]. \label{zeq1}
\end{eqnarray}
To find the optimal protocol,  the total evolution time is divided
 into N grids of width $\Delta t$ such that $V(t)$ can be
represented by constant potential $V_j$ in the $j^{\rm th}$
interval. Since a constant $V_j$ corresponds to a constant $u\equiv
u_j$ and $K \equiv K_j$,  a recursive solution for $z_k^j$ can be derived
\begin{eqnarray} 
z_k^j= iK_j^{-1} \tan[ k u_j \Delta t +\arctan[-i K_j z_k^{j-1}]].
\end{eqnarray}
Its solution together with the knowledge of the final ground state suffices to
 compute the wavefunction overlap  \cite{RC11}. This is the quantity that
acts as a cost function and whose optimization is achieved by  varying
the set of parameters $u_j$ and $K_j$ using Monte Carlo techniques
 \cite{RC11}. We note that finding the cost function implies the
calculation of the wavefunction overlap of the system for an
arbitrary set of parameters. Whereas this allows for the determination of the
optimal protocol, such a computation poses a challenge when
the system at hand is non-integrable and of moderate size. We also note that the optimal control of integrable systems is of interest in its own right, and has been suggested as a route to
universal quantum computation  \cite{LM14}.

Optimal driving protocols have also been applied to other
models such as the two-level system and the 1D Heisenberg spin chain
 \cite{Caneva09}. One important aspect of such studies constitutes the
relation of optimal-protocol design to the so called ``quantum speed
limit", associated with a fundamental  bound to the minimum time required
for transition between two quantum states to occur, as dictated by  Schr\"odinger dynamics.
Early results restricted to time-independent Hamiltonians  \cite{MT45,Fleming73,bhat1,Uhlmann92,ML98,Lloyd00,Busch08}
have recently been extended to arbitrary driven systems in a variety of forms  \cite{AA90,Pfeifer93,Caneva09,CM10,DL13,Taddei13,DEPH13,DL13b},
although the question as to whether these new bounds are tight and reachable remains unresolved.
With that caveat, the quantum speed limit  for isolated driven systems reads  \cite{DL13b}
\begin{eqnarray}
T_{\rm QSL}\geq{\rm max}\bigg\{\frac{\hbar\,}{\overline{E}},\frac{\hbar\,}{\overline{\Delta E}}\bigg\}\,\sin^2\mathcal{L}\left(\psi_0,\psi_T\right).
\end{eqnarray}
Here, the angle between the states $\psi_0$ and $\psi_T$ is measured by the Bures length,
\begin{eqnarray}
\mathcal{L}\left(\psi_0,\psi_T\right)= \arccos|\langle\psi_0|\psi_T\rangle|.
\end{eqnarray}
The first part of the bound, known as the Margolus-Levitin (ML)  \cite{ML98},
limits  the speed of evolution by the inverse of  mean energy,
$$\overline{E}=\frac{1}{T}\int_0^{T}\{\langle\psi_t|H(t)|\psi_t\rangle
-E_0(t)\} dt.$$
 The second part of the bound generalizes  to driven
systems the Mandelstam-Tamm (MT) time-energy uncertainty relation
 \cite{MT45,Fleming73,bhat1,AA90,Uhlmann92,Busch08}, where the
time-averaged squared root of the  energy variance reads
%\begin{eqnarray}
$$
\overline{\Delta E}=\frac{1}{T}\int_0^{T}
\sqrt{\langle\psi_t|H(t)^2|\psi_t\rangle-\langle\psi_t|H(t)|\psi_t\rangle^2}
dt.
$$
%\end{eqnarray}
Using a variant of this result  \cite{Pfeifer93}, it was shown in
Ref.\ \cite{Caneva09} that if the evolution time $T$
happens to be shorter than $T_{\rm QSL}$, the optimization
algorithms do not converge. The power of this conclusion arises from
the fact that it is independent of the protocol chosen for the
drive; hence it provides a completely general bound for the maximum
speed attainable via optimal control. Quantum speed
limits also determine the solution of the quantum brachistochrone
problem aimed at the preparation of a target state in a minimum time 
starting from a given initial state
 \cite{Carlini06,Carlini07,Hegerfeldt13,Mohseni14}.

\section{Two-rate dynamics}
\label{sau1}
In this section, we review the details of a third method, namely,
the suppression of defect density on passage through a quantum
critical point when two parameters of the Hamiltonian are
simultaneously varied in time  \cite{kstr1,kstr2}. The method
relies on reducing the available phase space for
excitation production using a second control parameter. Such a
reduction does not necessarily lead to perfect shortcut to adiabaticity;
however, the method has the advantage of relatively straightforward
experimental implementation.

To provide a simple demonstration of this method, we first consider
its application to integrable models such as the XY and the Kitaev
models  \cite{kstr1} . As discussed in Sec.\
\ref{secCD}, these models can be described in terms of non-interacting fermions
via a Jordan Wigner transformation \cite{dc1,ks1}. The Hamiltonian of
such non-interacting fermions in $d$-dimensions can be written as
$H=\sum_k \psi_{\bf k} ^{\dagger} H_{\bf k} \psi_{\bf k}$  where
$\psi_{\bf k}^{\dagger}= (c_{1 \bf k}^{\dagger},c_{2 \bf
k}^{\dagger})$ are Fermionic creation operators and $H_{\bf k}(t)$
is given by
\begin{eqnarray}
H_{\bf k}(t) = \tau_3 (\lambda_1(t) -b_{\bf k}) + \tau_1
\lambda_2(t) g_{\bf k}. \label{ham1}
\end{eqnarray}
Here $\tau_3$ and $\tau_1$ denote the usual Pauli matrices while $b_{\bf
k}$ and $g_{\bf k}$ are general functions of momenta, and
$\lambda_{1(2)}(t)= \lambda_0 \omega_{1(2)} t$ are time dependent
parameters driven with ramp rates $\omega_{1(2)}$. Note that in
contrast to the usual driving schemes leading to the scaling laws in Eq. (\ref{eq:scalinglaws}) where
only $\lambda_1$ is taken to be a function of time  \cite{pol1}, we
have chosen to vary the off-diagonal term $\lambda_2$ as well. The
instantaneous eigenvalues of the Hamiltonian is given by $E_{\bf
k}(t) = \pm \sqrt{(\lambda_1(t) -b_{\bf k})^2+(\lambda_2(t) g_{\bf
k})^2}$. We assume that as a result of the ramps, the system
reaches the critical point at $t=t_{0{\bf k}_0}= b_{{\bf
k}_0}/\omega_1$ and ${\bf k}= {\bf k}_0$ where $g_{{\bf k}_0}=0$.

\begin{figure}[tbp]
\begin{center}
\includegraphics[width=0.5\textwidth]{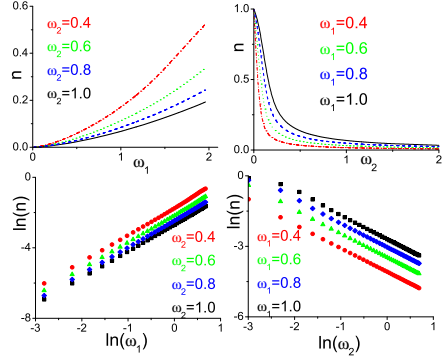}
\end{center}
\caption{(Color online) Top Panel: Plot of $n$ vs $\omega_1$ (left)
and $\omega_2$ (right) showing scaling of $n$. Bottom Panel: Plot of
$\ln(n)$ as a function of $\ln(\omega_1)$ (left) and $\ln(\omega_2)$
(right). All plots are computed using Eq. (\ref{exexp}) with $d=1$,
$b_{\bf k}=5-\cos(k)$, and $g(k)=\sin(k)$ so that $H$ represents 1D
XY model in a transverse field. The scaling regime, where Eq. 
(\ref{scaling2}) holds, occur for $\omega_2 \ge \omega_1^{3/2}/b_{\bf
k_0} =0.25 \omega_1^{3/2}.$ Taken from \cite{kstr1}.}\label{fig2}
\end{figure}

It turns out that the Schr\"odinger equation corresponding to $H_1(t)$
can be exactly solved. The key observation in this regard is that
$H_1$ can be written in terms of a set of new Pauli matrices
${\tilde \tau}_3$ and ${\tilde \tau}_1$ as follows,
\begin{eqnarray}
H_{\bf k}(t) &=& \lambda_{1{\bf k}} (t-t_{1{\bf k}}) {\tilde \tau_3}
+ \lambda_{2{\bf k}} {\tilde \tau}_1,  \label{ham2}
\end{eqnarray}
where $t_{1{\bf k}} = b_{\bf k} \omega_1/\lambda_{1{\bf k}}$. In the
above expression, the matrices ${\tilde \tau}_{1,3}$ can be
expressed in terms of $\tau_{1,3}$ as
\begin{eqnarray}
\lambda_{1{\bf k}} {\tilde \tau}_3 &=& \omega_1 \tau_3 + \omega_2
g_{\bf k} \tau_1,  \nonumber\\
\lambda_{2{\bf k}} {\tilde \tau}_1 &=& (\omega_1 t_{1{\bf k}}-b_{\bf
k}) \tau_3 + \omega_2 t_{1{\bf k}} g_{\bf k}  \tau_1. \label{exact2}
\end{eqnarray}
Further, the quantities $\lambda_{1{\bf k}}$ and $\lambda_{2{\bf k}}$ are
obtained by diagonalizing Eq. (\ref{exact2})  \cite{kstr1}:
$\lambda_{1{\bf k}}= [\omega_1^2+\omega_2^2 g_{\bf k}^2]^{1/2}$ and
$\lambda_{2{\bf k}}= t_{1{\bf k}} [(\omega_2-b_{\bf k}/t_{1{\bf
k}})^2 +\omega_2^2 g_{\bf k}^2]^{1/2}$.

The above transformation shifts the entire time dependence of
$H_{\bf k}(t)$ to diagonal terms and reduces the corresponding
Schr\"odinger equation to a set Landau-Zener problem (one for each
mode ${\bf k}$)  \cite{lz1}. Using the results of Refs.\
 \cite{lz1,vitanov1}, one can thus simply read off the
probability of excitation (defect) production for any ${\bf k}$ to
be $p_{\bf k} = \exp[- \pi \lambda_{2{\bf k}}^2/\lambda_{1{\bf k}}]$
which leads to the defect density
\begin{eqnarray}
n  &=&  \int d^dk /(2 \pi)^d e^{- \pi \omega_{2}^2 b_{\bf k}^2
g_{\bf k}^2/(\omega_1^2+\omega_2^2 g_{\bf k}^2)^{3/2}}.
\label{exexp}
\end{eqnarray}
For $\omega_{1,2}/b_{\bf k_0}^2 \ll 1 $ and $\omega_2/\omega_1 \ge
\sqrt{\omega_1}/b_{\bf k_0}$, $p_{\bf k}$ is appreciable around
${\bf k}= {\bf k}_0$. Thus in this regime, one can replace $p_{\bf
k}$ by its value around ${\bf k_0}$ leading $p_{\bf k}=
e^{-c\,\omega_2^2 k^{2}/\omega_1^3}$, with $c= \pi b_{{\bf k}_0}^2$.
A simple rescaling of $p_{\bf k}$, $k' = k \omega_2/\omega_1^{3/2}$,
in this regime then leads to Eq. (\ref{exexp}),
\begin{eqnarray}
n \sim \omega_1^{3d/2} \omega_2^{-d}, \label{scaling2}
\end{eqnarray}
which demonstrates  the suppression of  the density of excitations with increasing
$\omega_2$. Note that the validity of the scaling relations does not
constrain $\omega_2/\omega_1$ to small values; thus one can
efficiently suppress defects by tuning $\omega_2$ for a suitably
chosen $\omega_1$. A plot of $n$ computed from Eq. (\ref{exexp}) with
$d=1$, $b_{\bf k}= 5-\cos(k)$, and $g_{\bf k}= \sin(k)$ (chosen so
that the model conforms to 1D XY model in a transverse field) is
shown in top panels of Fig.\ \ref{fig1} as a function of the rates
$\omega_1$ and $\omega_2$. The plot clearly demonstrates that $n$ is
a decreasing function of $\omega_2$.

Eq. (\ref{scaling2}) indicates the existence of two separate regimes
$n$ behaves qualitatively differently with $\omega_1$ with
$\omega_2=\omega_1^{r}$. In one regime, where $r\ge 3/2$, $n$
increases $\omega_1$ while it decreases with $\omega_1$ if $r<3/2$
The crossover between these regimes occurs for $\omega_2 =
\omega_1^{r^{\ast}}$ with $r^{\ast}=3/2$ for any $d$. This crossover
is indicated in Fig.\ \ref{fig2}, where $n$ is plotted as function
of $\omega_1$ with $\omega_2=\omega_1^r$. We note here that at
$r=r^{\ast}$, $n$ becomes independent of $\omega_1$ and $\omega_2$.

\begin{figure}[tbp]
\begin{center}
\includegraphics[width=0.4\textwidth]{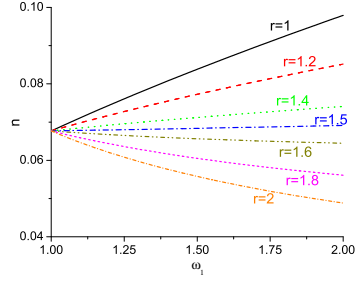}
\end{center}
\caption{Tailoring  excitation formation by two-rate dynamics.
Dependence of $n$ on $\omega_1$ with $\omega_2=
\omega_1^r$ showing the crossover between regimes with increasing
and decreasing $n$ as a function of $\omega_1$. All parameters are
the same as in Fig.\ \ref{fig2}.Taken from \cite{kstr1}.} \label{fig3}
\end{figure}

It turns out that it is relatively straightforward to generalize
these concepts for arbitrary time dependent Hamiltonians. To see
this, let us consider a generic Hamiltonian with two tunable
parameters which are varied with rates $\omega_1$ and $\omega_2$.
The first parameter $\lambda(t)$ controls its distance from a
quantum critical point at $\lambda=\lambda_c \ne 0$; this
necessitates that the instantaneous energy gap near the critical
point varies as $\Delta({\bf k}={\bf k}_0;\lambda) \simeq
|\lambda(t)|^{z\nu} = |\omega_1 t -\lambda_c|^{z\nu \alpha}$, where
$\alpha$ is a positive exponent and $\alpha=1$ denotes linear drive
protocol. The second parameter, $c(t)$, controls the dispersion of
the quasiparticles at the critical point so that
$\Delta(k,\lambda_c) \simeq c(t) k^z = |\omega_2 t|^{\beta} k^z$.

To estimate the defect density generated during such a drive, we
first estimate the time spent by the system in the impulse region
where defect production occurs (for small $\omega_1$ it is also the
critical region). For this, we use the well-known Landau criterion
which states that a quantum system subjected to a drive is in the
impulse region if  $d \Delta/dt \simeq \Delta^2$ \,  \cite{lz1,pol1}.
Substituting the expression for $\Delta({\bf k}_0,\lambda(t))$ in
this relation, one obtains an expression for $T$, the time spent by
the system in the impulse region, as
\begin{eqnarray}
|T-T_0| \simeq \omega_1^{-\alpha z\nu /(\alpha z \nu +1)},
\label{t0exp}
\end{eqnarray}
where $T_0= \lambda_c/\omega_1$ is the time at which the system
reaches the critical point. Substituting the expression for $T$ in
that for $\Delta({\bf k}_0,\lambda)$, one finds that in the impulse
region, the instantaneous energy gap behaves as
\begin{eqnarray}
\Delta({\bf k}_0;\lambda) \simeq \omega_1^{\alpha z\nu /(\alpha z
\nu +1)}. \label{enbehav}
\end{eqnarray}
Next, one notes that the defects or excitations are typically
produced in a phase space $\Omega \sim k^d$ around the critical
mode. During the time $T$ that the system spends in the impulse
region, these momentum modes satisfy  \cite{comment3}
\begin{eqnarray}
k &\simeq& |\omega_2 T_0|^{-\beta/z} \Delta({\bf
k}_0,\lambda(T))^{1/z}  \label{komegarel}
\end{eqnarray}
Using Eqs. (\ref{t0exp}), (\ref{komegarel}) and (\ref{enbehav}), one
finally gets
\begin{eqnarray}
n \sim \Omega \sim k^d \simeq \omega_2^{-\beta d/z}
\omega_1^{\left(\frac{\alpha \nu}{\alpha z \nu +1} +
\frac{\beta}{z}\right) d}.  \label{nfinal}
\end{eqnarray}
which generalizes Eq. (\ref{scaling2}). The present analysis shows
that the suppression of $n$ with increasing $\omega_2$ occurs due to
the reduction of available momentum modes for quasiparticle
excitations at any given energy $\Delta(k;\lambda)$; thus the role
of the drive protocol changing $c(t)$ is to reduce the available
phase space for defect production. Analyzing Eq. (\ref{nfinal}), one
finds ${r^\ast}= 1+ \alpha z \nu/(\beta(\alpha z \nu +1))$ which
reduces to the condition $r^{\ast}=3/2$ derived earlier for
$\alpha=\beta=z=\nu=1$. We note in passing that Eq. (\ref{nfinal})
also constitute a generalization of Kibble-Zurek scaling laws for
two-rate protocols.

As discussed in Ref.\ \cite{kstr1}, there are several concrete
models where the present method may be applied. However, it is
perhaps more interesting to note that  quantum systems near a
phase transitions can often be described by a Landau-Ginzburg action which
has the generic form
\begin{eqnarray}
S &=& \int d^dr dt  \psi^{\ast} [ - \partial_t^2 + c_1 \sum_{i=1,d}
\partial_{x_i}^{2z} + (r-r_c) - u |\psi|^2 ]\psi, \nonumber
\end{eqnarray}
where $r$ controls the distance to criticality while $c_1$ controls
the quasiparticle dispersion at criticality. Thus, if $r$ and $c_1$
are tuned as functions of time with rates $\omega_1$ and $\omega_2$,
one expects phase-space suppression leading to defect reduction as a
function of $\omega_2$. This indicates that the suppression
discussed above is of general nature. However, it is to be observed
that $r$ and $c_1$ needs to be obtained from the microscopic
parameters of the system action; thus whereas defect reduction
occurs generically if $c$ is increased, one still needs to specify
the relation between the effective parameter $c$ to microscopic
parameters of $H$ which can be experimentally tuned. This could be
difficult for generic actions and specially so, for strongly
interacting systems. Some progress in this direction has recently
been made  \cite{kstr2,comment4,Ian1,coldatomMajorana}.

\section{Conclusion}
\label{conc}

In this review, we have discussed several methods for tuning the
excitation production in a closed quantum system during its passage
through a quantum critical point. Among a broad variety of
alternative routes to achieve this reduction, we have presented
three techniques in detail. The first one, discussed in Sec.\ref{secCD},
involves engineering an additional term $H_1$ for the system
Hamiltonian $H_0$, such that the dynamics generated by $H(t)=H_0(t)+H_1(t)$
follows the adiabatic manifold of $H_0$.
%designed  such that the adiabatic approximation
%under $H_0$  becomes the exact
%solution of the many-body time dependent Schr\"odinger equation with
%$H(t)=H_0(t)+H_1(t)$. 
Determining the auxiliary term requires access
to the spectral properties of the system $H_0$ and its
implementation might involve non-local multiple-body interactions.
The strength of this method lies in the possibility of reducing
excitation  formation via approximate construction of $H_1$ under
given resources. The second method involves the determination of an
optimal time-dependence of the system Hamiltonian $H_0(t)$ using
optimal control to maximize the overlap between the time evolving
state and the target state, as discussed in Sec.\ \ref{octh1}. This
method is mathematically rigorous; however, its implementation
requires knowledge of the time-dependent many-body state of the
system during the evolution which limits its applicability to
moderate system sizes in non-integrable models. Finally, the
two-rate protocol discussed in Sec.\ \ref{sau1} exploits a
two-parameter tuning of the system Hamiltonian. One of these
parameters reduces  the phase space available for  excitation
formation, and consequently,  it suppresses defect production. While
the method is not optimal, it allows in principle for an easy
implementation in many-systems since only one additional parameter
of the system Hamiltonian is to be tuned. However, the
identification of the second drive parameter is system specific and
at present, it has been theoretically tested for only a handful of
non-integrable many-body systems.

It is our hope that the ideas summarized in this review contribute
to deepen our understanding of the far-from-equilibrium dynamics of
isolated quantum systems and related research areas.  New
theoretical and experimental developments can be expected pursuing
applications of controlled quantum critical dynamics in the field of
quantum simulation  \cite{CZ12}, thermalization of isolated quantum
systems  \cite{EFG14} and work fluctuations in finite-time
thermodynamics  \cite{CHT11}. The implications of these techniques
in the design
 of new protocols to assist and speed up quantum methods
for optimization   \cite{Boixo14} constitute another research direction worth exploring.

\begin{acknowledgement}
We acknowledge  B. Damski, S. Deffner, A. Dutta,  S. Montangero, B. Peropadre, H.
Saberi, J. D. Sau,  D. Sen, and F. Setiawan for useful discussions
and suggestions.  AdC further thanks N. Guler for hospitality during
the completion of the manuscript.
\end{acknowledgement}


\begin{thebibliography}{99}
\bibitem{comment3} In principle, one should evaluate $c(t)$ at $t=T$
instead of $t=T_0$ as we have done; however, it is easy to see from
the expression of $T$ that this difference leads to subleading
corrections which are irrelevant for the present purpose.
\end{thebibliography}
\end{document}